\documentclass[12pt]{article}
\usepackage{epsf}
\title{ {\bf
$t\rightarrow b W h^0$ and $t\rightarrow b W A^0$ decays and possible CP
violating effects.}}
\author{\vspace{1cm}\\
        {\bf E. O. Iltan}
        \thanks{E-mail address:
        eiltan@heraklit.physics.metu.edu.tr}
 \\
        Physics Department, Middle East Technical University \\
        Ankara, Turkey\\}

\date{}

\begin{document}
\setlength{\baselineskip}{24pt}
\maketitle
\setlength{\baselineskip}{7mm}
\begin{abstract}
We study the charged $t\rightarrow b W h^0$ and $t\rightarrow b W A^0$ 
decays in the framework of the general two Higgs doublet model, so called 
model III and beyond. Here, we take the Yukawa couplings complex and
introduce a new complex parameter due to the physics beyond the model III, 
to switch on the CP violating effects. We predict the branching ratios 
as $BR (t\rightarrow b W h^0) \sim 10^{-6}$ and 
$BR (t\rightarrow b W A^0) \sim 10^{-8}$. Furthermore, we observe a 
measurable CP asymmetry, at the order of $10^{-2}$, for both decays.
\end{abstract} 
\thispagestyle{empty}
\newpage
\setcounter{page}{1}
\section{Introduction}
Because of its large mass, the top quark has rich decay products and this 
opens a new window to test the standard model (SM) and to get some clues 
about the new physics, beyond. In the literature there are various studies 
in the SM and beyond \cite{Mahlon}-\cite{Alterelli}.The rare flavor changing 
transitions $t\rightarrow c g(\gamma, Z)$ have been studied in 
\cite{Liu,Eilam3}, $t\rightarrow c H^0$ in \cite{Eilam, Eilam3, 
Mele, Tao, IltanH0} and $t\rightarrow c l_1 l_2$ in \cite{Iltanl1l2}. 
The SM predictions of the branching ratio ($BR$) of the process 
$t\rightarrow c g(\gamma, Z)$ is  $4 \times 10^{-11}\, (5 \times 10^{-13},\, 
1.3 \times 10^{-13}\,)$ \cite{Eilam}, and  $t\rightarrow c H^0$ is at the 
order of the magnitude of $10^{-14}-10^{-13}$, in the SM \cite{Mele}, which
are not measurable quantities even at the highest luminosity accelerators.
Possible new physics effects are the candidates for the enhancement of the
$BR$'s of the above processes. $t\rightarrow c H^0$ and 
$t\rightarrow c l_1 l_2$ decays have been analysed in \cite{IltanH0} and 
\cite{Iltanl1l2}, in the framework of the general two Higgs doublet model 
(model III). In these studies  it has been observed that there could be a 
strong enhancement in the $BR$,  almost seven orders larger 
compared to the one in the SM for the decay $t\rightarrow c H^0$; a
measurable $BR$, at the order of the magnitude of $10^{-8}-10^{-7}$ 
for the decay $t\rightarrow c l_1 l_2$. In \cite{Yue} $t\rightarrow c V V$
decay has been analysed in the topcolor assisted technicolor theory.

The charged $t\rightarrow b$ transitions exist in the SM model and have been
studied in the literature extensively. The top decay $t\rightarrow b W $ has
been analysed (see \cite{Denner} and references therein) in the two Higgs
doublet model and  $t\rightarrow b W Z$ decay has been studied in 
\cite{Alterelli}.  

The present work is devoted to the analysis of the charged 
$t\rightarrow b \,W \,  h^0$ and $t\rightarrow b \,W \,  A^0$ decays in 
the framework of the general two Higgs doublet model (model III). This 
decay occurs in the tree level with the extended Higgs sector since the
scalar bosons $h^0$ and $A^0$ exist in the new sector. We study the $BR$ of
the above decays and obtain a measurable quantities, at the order of the
magnitude of $10^{-6}$ and  $10^{-8}$, respectively. Furthermore, we search 
the possible CP violating effects. To obtain a nonzero CP asymmetry
$A_{CP}$ we take Yukawa coupling for $b\,h^0\,(A^0)\, b$ transition complex
and introduce a new complex parameter, where its complexity comes from some 
type of radiative corrections, due to the model beyond the model III 
(see section II). We obtain a measurable $A_{CP}$, at the order of the 
magnitude of $10^{-2}$ and observe that these physical quantities can
give valuable information about physics beyond the SM, the free parameters
existing in these models.     

The paper is organized as follows:
In Section 2, we present the $BR$ and $A_{CP}$ of the decay $t\rightarrow b
\,W\, h^0 (A^0)$ in the framework of model III. Section 3 is devoted to 
discussion and our conclusions.
\section{$t\rightarrow b W h^0$ and $t\rightarrow b W A^0$ decays with
possible CP violating effects.} 
If one respects the current mass values of $h^0 (A^0)$, namely $m_{h^0}\sim
85\, GeV$ ($m_{A^0}\sim 90\, GeV$), the charged $t\rightarrow b W h^0 (A^0)$ 
is kinematically possible and does not exist in the SM model. With the 
minimal extension of the  Higgs sector the CP odd new Higgs scalar $A^0$ 
arises and the $t\rightarrow b W A^0$ decay in the tree level is permitted. 
In this model, $t\rightarrow b W h^0$ decay is possible in the tree
level, where $h^0$ is the new CP even Higgs scalar and, in general, it 
mixes with the SM one, $H^0$. In this section, we study the $BR$ in the 
general two Higgs doublet model, so called model III and the
possible CP violating asymmetry, beyond. 

The $t\rightarrow b W h^0 (A^0)$ decay is created by the charged 
$t\rightarrow b W$ process and the neutral  $t\rightarrow t^* h^0 (A^0)$ or 
$b^* \rightarrow b h^0 (A^0)$ processes, which are controlled by the Yukawa 
interaction  
\begin{eqnarray}
{\cal{L}}_{Y}=\eta^{U}_{ij} \bar{Q}_{i L} \tilde{\phi_{1}} U_{j R}+
\eta^{D}_{ij} \bar{Q}_{i L} \phi_{1} D_{j R}+
\xi^{U\,\dagger}_{ij} \bar{Q}_{i L} \tilde{\phi_{2}} U_{j R}+
\xi^{D}_{ij} \bar{Q}_{i L} \phi_{2} D_{j R} + h.c. \,\,\, ,
\label{lagrangian}
\end{eqnarray}
where $L$ and $R$ denote chiral projections $L(R)=1/2(1\mp \gamma_5)$,
$\phi_{i}$ for $i=1,2$, are the two scalar doublets, 
$\bar{Q}_{i L}$ are left handed quark doublets, $U_{j R} (D_{j R})$ are 
right handed up (down) quark singlets, with  family indices $i,j$. The 
Yukawa matrices $\eta^{U,D}_{ij}$ and $\xi^{U,D}_{ij}$ have in general 
complex entries. By considering the gauge and $CP$ invariant Higgs 
potential which spontaneously breaks  $SU(2)\times U(1)$ down to $U(1)$ 
as
\begin{eqnarray}
V(\phi_1, \phi_2 )&=&c_1 (\phi_1^+ \phi_1-v^2/2)^2+
c_2 (\phi_2^+ \phi_2)^2 \nonumber \\ &+& +
c_3 [(\phi_1^+ \phi_1-v^2/2)+ \phi_2^+ \phi_2]^2
+ c_4 [(\phi_1^+ \phi_1) (\phi_2^+ \phi_2)-(\phi_1^+ \phi_2)(\phi_2^+ \phi_1)]
\nonumber \\ &+& 
c_5 [Re(\phi_1^+ \phi_2)]^2 +
c_{6} [Im(\phi_1^+ \phi_2)]^2 
+c_{7}\,\, .
\label{potential}
\end{eqnarray}
and choosing the parametrization for $\phi_{1}$ and $\phi_{2}$ as
\begin{eqnarray}
\phi_{1}=\frac{1}{\sqrt{2}}\left[\left(\begin{array}{c c} 
0\\v+H^{0}\end{array}\right)\; + \left(\begin{array}{c c} 
\sqrt{2} \chi^{+}\\ i \chi^{0}\end{array}\right) \right]\, ; 
\phi_{2}=\frac{1}{\sqrt{2}}\left(\begin{array}{c c} 
\sqrt{2} H^{+}\\ H_1+i H_2 \end{array}\right) \,\, .
\label{choice}
\end{eqnarray}
with the vacuum expectation values,  
\begin{eqnarray}
<\phi_{1}>=\frac{1}{\sqrt{2}}\left(\begin{array}{c c} 
0\\v\end{array}\right) \,  \, ; 
<\phi_{2}>=0 \,\, ,
\label{choice2}
\end{eqnarray}
the $H_1$ and $H_2$ becomes the mass eigenstates $h^0$ and $A^0$ respectively 
since no mixing occurs between two CP-even neutral bosons $H^0$ and $h^0$, 
in tree level. This scenerio permits one  to collect SM particles in the 
first doublet and new particles in the second one. Furthermore the Flavor 
Changing (FC) interaction can be obtained as 
\begin{eqnarray}
{\cal{L}}_{Y,FC}=
\xi^{U\,\dagger}_{ij} \bar{Q}_{i L} \tilde{\phi_{2}} U_{j R}+
\xi^{D}_{ij} \bar{Q}_{i L} \phi_{2} D_{j R} + h.c. \,\, ,
\label{lagrangianFC}
\end{eqnarray}
with the couplings  $\xi^{U,D}$ for the FC charged interactions 
\begin{eqnarray}
\xi^{U}_{ch}&=& \xi^U_{N} \,\, V_{CKM} \nonumber \,\, ,\\
\xi^{D}_{ch}&=& V_{CKM} \,\, \xi^D_{N} \,\, ,
\label{ksi1} 
\end{eqnarray}
where $\xi^{U,D}_{N}$ is defined by the expression 
\begin{eqnarray}
\xi^{U (D)}_{N}=(V_{R (L)}^{U (D)})^{-1} \xi^{U,(D)} V_{L(R)}^{U (D)}\,\, .
\label{ksineut}
\end{eqnarray}
Notice that the index "N" in $\xi^{U,D}_{N}$ denotes the word "neutral". 

\begin{figure}[htb]
\vskip 1.5truein
\centering
\epsfxsize=6.8in
\leavevmode\epsffile{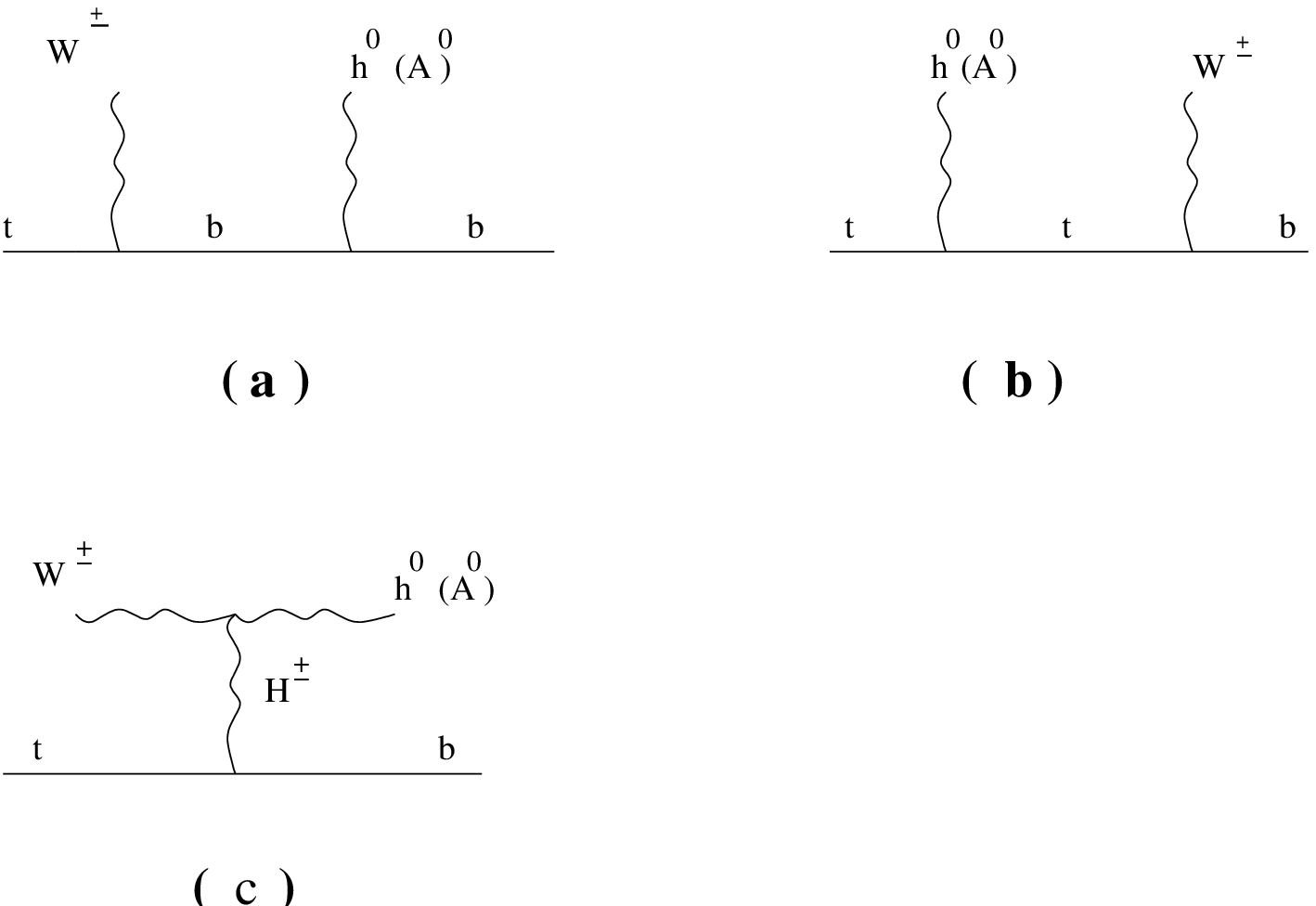}
\vskip 1.0truein
\caption[]{The diagrams contribute to the decay $t\rightarrow b W h^0
(A^0)$.}
\label{fig1}
\end{figure}

Using the relevant diagrams for the $t\rightarrow b W h^0 (A^0)$ decay 
which are given in Fig \ref{fig1} and taking into account only the real 
Yukawa couplings $\xi^D_{N,bb},\,\xi^U_{N,tt}$, the matrix element square 
$|M|^2 (h^0)$ ($|M|^2 (A^0)$) reads 
\begin{eqnarray}
|M|^2_{h^0\, (A^0)} (p1,p_b,k,q) = \xi^D_{N,bb}\,\xi^U_{N,tt}\,
f_1 (h^0 \, (A^0))+ (\xi^D_{N,bb})^2\,\,f_2 (h^0 \, (A^0))+(\xi^U_{N,tt})^2
\,f_3 (h^0 \, (A^0)) 
\label{M2h0A0}
\end{eqnarray}
where
\begin{eqnarray}
f_1 (h^0)&=& 16\, |V_{tb}|^2\,m_b\, m_t\, \Bigg( m_W^2 \Big( s_2^2 (h^0)-
s_1^2 (h^0)+ 2\, s_1 (h^0) \, s_2 (h^0) \Big) x_{h^0}
\nonumber \\ &+&
2\, s_1 (h^0)\, \Big( (s_1 (h^0)-s_2 (h^0))\, k.(p_1-p_b)+2\, s_1 (h^0)\,
p_1.p_b \Big) 
\nonumber \\ &+&
\frac{1}{m_W^2}\, \Big( -(s_2^2 (h^0)+ 2\, s_1^2 (h^0)+ 2\, s_1 (h^0) \, 
s_2 (h^0) )\, (k.q)^2+ 2\,s_1 (h^0)\, (2\,s_1 (h^0)\nonumber \\ 
&+& s_2 (h^0))\, k.q\, q.(p_1-p_b) + 8\,s_1^2 (h^0)\, p_1.q\, p_b.q \Big )  
\Bigg) 
\, , \nonumber \\ 
f_2 (h^0)&=& 8\, |V_{tb}|^2 \, \Bigg(-m_W^2 \Big( s_2 (h^0)+
s_1 (h^0)\Big)^2 \,x_{h^0} \,p_1.p_b 
\nonumber \\ &+& \!\!\!\!
\frac{1}{m_W^2}\, k.q\,\Big( (s_2 (h^0)+2\, s_1 (h^0) )\, 
\nonumber \\ &\times& 
(s_2 (h^0)\, k.q\,p_1.p_b+2\,s_1 (h^0)\, k.p_b\, q.p_1)-
2\, s_1 (h^0)\,s_2 (h^0)\, k.p_1 \, p_b.q \Big) 
\nonumber \\ &+&
2\,s_1^2 (h^0)\, (k.p_1\, k.p_b - x_{h^0}\,q.p_1\, q.p_b ) \Bigg )
\, ,\nonumber \\ 
f_3 (h^0)&=&\!\!\!\! 8\, |V_{tb}|^2\,\Bigg(-m_W^2 \Big( 4\, s_1 (h^0)\, 
(s_1 (h^0)-s_2 (h^0))\, x_t\, k.p_b+ \Big( (s_1 (h^0)+s_2 (h^0))^2\, x_{h^0} 
\nonumber \\ &-& 
4\,s_1^2 (h^0)\,x_t \Big)\, p_1.p_b \Big) 
\nonumber \\ &+&
\frac{1}{m_W^2}\, k.q\,\Big( s_2 (h^0)\,(s_2 (h^0)+ 2\, s_1 (h^0) )\, k.q\,
p_1.p_b-2\, s_1 (h^0)\,s_2 (h^0)\, q.p_1\, k.p_b +2\,s_1 (h^0) 
\nonumber \\ &\times& (2\,s_1 (h^0)+s_2 (h^0))\, k.p_1\,q.p_b \Big) 
\nonumber \\ &+& \!\!\!\!
2\,s_1 (h^0)\, \Big( s_1 (h^0)\,k.p_1\, k.p_b - \Big( 2\,(2\, s_1 (h^0)+
s_2 (h^0) )\,x_t\,k.q+s_1 (h^0)\,(x_{h^0}-4\,x_t)\, q.p_1 \Big) \,q.p_b 
\Big) \Bigg ) \, ,
\nonumber \\
f_1 (A^0) &=& 16\, |V_{tb}|^2\,m_b\, m_t\, \Bigg( m_W^2 \Big( s_2^2 (A^0)+
s_1^2 (A^0) \Big)\, x_{A^0}+ \frac{1}{m_W^2}\Big( 2\, s^2_1 (A^0)-s^2_2 (A^0) 
\Big)\, (k.q)^2\, \Bigg) 
\nonumber \, ,\\ 
f_2 (A^0)&=& 8\, |V_{tb}|^2\, \Bigg( -m_W^2 \Big( s_2 (A^0)+
s_1 (A^0)\Big)^2 \,x_{A^0}\, p_1.p_b
\nonumber \\ &+&
\frac{1}{m_W^2}\, k.q\,\Big( (s_2 (A^0)+ 2\, s_1 (A^0) )\, (s_2 (A^0)\, k.q\,
p_1.p_b+2\,s_1 (A^0)\, k.p_b\, p_1.q )\nonumber \\ &-& 
2\, s_1 (A^0)\,s_2 (A^0)\, k.p_1\, p_b.q \Big) +
2\,s_1^2 (A^0)\, (k.p_1\, k.p_b - x_{A^0}\,q.p_1\, q.p_b ) \Bigg )
\nonumber \, , \\ 
f_3 (A^0)&=& 8\, |V_{tb}|^2\,\Bigg(-m_W^2 \Big( s_1 (h^0)
-s_2 (h^0) \Big)^2\, x_{A^0}\, p_1.p_b
\nonumber \\ &+&
\frac{1}{m_W^2}\, k.q\,\Big( s_2 (A^0)\, (s_2 (h^0)- 2\, s_1 (h^0) )\, k.q\,
p_1.p_b+2\,s_1 (A^0)\, (2\,s_1 (A^0)-s_2 (A^0))\, k.p_1\,p_b.q )
\nonumber \\ &+&
2\,s_1 (A^0)\,s_2 (A^0)\,k.p_b\, p_1.q \Big)+
2\,s^2_1 (A^0)\, \Big( k.p_1\, k.p_b- x_{A^0} \, p_1.q\, p_b.q \Big) \Bigg ) 
\, .
\label{f123}
\end{eqnarray}
Here the functions $s_{1(2,3)}(h^0 (A^0))$ are 
\begin{eqnarray}
s_{1}(h^0)&=& -\frac{g_W}{4\,m_W^2\,(1+x_t-2\,\frac{p_1.q}{m_W^2})} 
\, ,\nonumber \\ 
s_{2}(h^0)&=& \frac{g_W}{2\,m_W^2\,(1+x_{h^0}-y_t-2\,\frac{k.q}{m_W^2})} 
\, ,\nonumber \\ 
s_{1 (2)}(A^0)&=& (-)\, s_{1 (2)}(h^0\rightarrow A^0) \, ,
\label{s1s2}
\end{eqnarray}
with weak coupling constant $g_W$, $x_{h^0 (A^0)}=\frac{m^2_{h^0 (A^0)}}
{m^2_W}$, $x_t=\frac{m^2_t}{m^2_W}$ and $y_t=\frac{m^2_{H^{\pm}}}{m^2_W}$ 
and $p_1$, $p_b$, $q$ and $k$ are four momentum of $t$ quark, $b$ quark, 
W boson and Higgs scalar $h^0 (A^0)$, respectively. 

Finally, using the well known expression defined in the $t$ quark rest frame
\begin{eqnarray}
d\Gamma_{h^0\, (A^0)}&=&\frac{(2\pi)^4}{12\,m_t}\, \delta^{(4)}(p_1-p_b-k-q)\, 
\frac{d^3 p_b}{(2\pi)^3\,2\,E_b}\,\frac{d^3 q}{(2\pi)^3\,2\,E_W} \, 
\frac{d^3 k}{(2\pi)^3\,2\,E_{h^0 (A^0)}} 
\nonumber \\ &\times& |M|^2_{h^0\, (A^0)} (p_1,p_b,k,q)
\label{DecWid} 
\end{eqnarray}
and the total decay width $\Gamma_T \sim \Gamma (t\rightarrow b W)$ as 
$\Gamma_T=1.55\, GeV $, we get the $BR$ for the decay 
$t\rightarrow b W h^0 (A^0)$.  

Now, we would like to study a possible CP violating effects, which can give
comprehensive information about the free parameters of the model used. For
the process under consideration, the CP violation can be obtained by choosing
the complex Yukawa couplings in general, namely, taking the parametrizations 
\begin{eqnarray}
\xi^{U}_{N,tt}=|\xi^{U}_{N,tt}|\, e^{i \theta_{tt}}
\nonumber \, , \\
\xi^{D}_{N,bb}=|\xi^{D}_{N,bb}|\, e^{i \theta_{tb}} \, .
\label{xi}
\end{eqnarray}
However, this choice is not enough to get non-zero $A_{CP}$ 
\begin{eqnarray}
A_{CP}=\frac{\Gamma - \bar{\Gamma}}{\Gamma + \bar{\Gamma}}
\label{ACP1}
\end{eqnarray}
where $\bar{\Gamma}$ is the decay width for the CP conjugate process. 
This forces one to go beyond the model III and try to obtain a new
complex quantity so that its complexity does not come from the Yukawa
couplings but from some radiative corrections. Under the light of this
discussion, we introduce an additional complex correction $\chi$ to 
$b\rightarrow b$ transition, which may come from the new 
model beyond the model III as 
\begin{eqnarray}
(\xi^D_{N,bb}+\xi^{D *}_{N,bb}) \!\!\! &+& \!\!\!(\xi^D_{N,bb}-
\xi^{D *}_{N,bb}) \gamma_5 + \chi 
\nonumber \\ &\mbox{and}&  \nonumber \\
(\xi^{D *}_{N,bb}-\xi^D_{N,bb}) \!\!\! &-&\!\!\!(\xi^D_{N,bb}+
\xi^{D *}_{N,bb}) \gamma_5 + \chi \gamma_5 \nonumber\, ,
\end{eqnarray}
Here we take the magnitude of $\chi$ at most $|\chi|\sim 10^{-2}$, which 
is more than one order smaller compared to the vertex due to model III.
In this case, we take the correction to the $t\rightarrow t$ transition 
small since the strength of $t\rightarrow t$ transition is weaker compared 
to strength of the $b\rightarrow b$ transition, with respect to our choice 
(see Discussion section). 

At this stage, we introduce a model beyond the model III as follows: The 
multi Higgs doublet model which contains more than two Higgs doublets in the 
Higgs sector can be one of the candidate. The choice of three Higgs doublets 
brings new Yukawa couplings which are responsible with the interactions 
between new Higgs particles and the fermions. The Yukawa lagrangian in three 
Higgs doublet model (3HDM) reads
\begin{eqnarray}
{\cal{L}}_{Y}&=&\eta^{U}_{ij} \bar{Q}_{i L} \tilde{\phi_{1}} U_{j R}+
\eta^{D}_{ij} \bar{Q}_{i L} \phi_{1} D_{j R}+
\xi^{U\,\dagger}_{ij} \bar{Q}_{i L} \tilde{\phi_{2}} U_{j R}+
\xi^{D}_{ij} \bar{Q}_{i L} \phi_{2} D_{j R} + 
\rho^{U}_{ij} \bar{Q}_{i L} \tilde{\phi_{3}} U_{j R}
\nonumber \\ &+&
\rho^{D}_{ij} \bar{Q}_{i L} \phi_{3} D_{j R} + h.c. \,\,\, ,
\label{lagrangian3H}
\end{eqnarray}
where $\rho^{U (D)}_{ij}$ is the new coupling and $\phi_{3}$ can be chosen 
as 
\begin{eqnarray}
\phi_{3}=\frac{1}{\sqrt{2}}\left(\begin{array}{c c} 
\sqrt{2} F^{+}\\ H_3+i H_4 \end{array}\right) \,\, ,
\label{choice3H}
\end{eqnarray}
with vanishing vacuum expectation value. The fields $F^{+}$ and $H_3\, 
(H_4)$ represent the new charged and CP even (odd) Higgs particles, 
respectively. Notice that the other Yukawa couplings and Higgs particles in 
eq. (\ref{lagrangian3H}) are the ones existing in the model III.  Now, 
we choose the additional Yukawa couplings $\rho^{U (D)}_{ij}$ real and take 
into account the radiative corrections to the  $b\rightarrow b$ transition 
which comes from the contributions of third Higgs doublet for the decay under 
consideration. Here the complexity of the parameter should come from the 
radiative corrections but not from the new Yukawa couplings. We can take 
this complex contribution as a source for the additional part $\chi$. 
Since the number of free parameters, namely masses of new Higgs particles 
$m_{F^{\pm}},\, m_{H_3}, \, m_{H_4}$ and the new Yukawa couplings 
$\rho^{U (D)}_{ij}$, increases, there arises a difficulty to restrict them.
However, the overall uncertainity coming from these free parameters lies in 
the contribution $\chi$ and it can be overcome by the possible future 
measurement of the CP violation for our process.

Finally, by using the definition 
\begin{eqnarray}
A^{h^0 (A^0)}_{CP}(E_W, E_b)= \frac{\frac{d^2 \Gamma (t\rightarrow b W h^0 
(A^0))} {d E_b d E_W} - \frac{d^2\Gamma (\bar{t}\rightarrow 
\bar{b} \bar{W} h^0 (A^0)}
{d E_b d E_W}} {\frac{d^2 \Gamma (t\rightarrow b W h^0 (A^0))}
{d E_b d E_W} +\frac{d^2\Gamma (\bar{t}\rightarrow \bar{b} \bar{W} h^0 
(A^0)} {d E_b d E_W}}
\label{cpvio}
\end{eqnarray}
we obtain the differential $A_{CP}(E_W, E_b)$ for the process 
$t\rightarrow b W h^0 (A^0)$ as
\begin{eqnarray}
A^{h^0 (A^0)}_{CP}(E_W, E_b)&=& |\bar{\xi}^D_{N,bb}|\, |\chi| \, sin\theta_{bb} \, 
sin\theta_{\chi}\frac{\Phi^{h^0 (A^0)}}{D^{h^0 (A^0)}} \, ,\nonumber \\
\label{ACP2}
\end{eqnarray}
where
\begin{eqnarray}
\Phi^{h^0}&=& 4\,m_t\,s_1(h^0)\, |V_{tb}|^2\, \Bigg( 4
\Big(2\,E_W^2\,m_t\,s_1(h^0)\, (x_t-2)-E_b^2\, (2\,E_W+m_t)\,s_2 (h^0)\,
(1+x_t) \nonumber \\ &+&
E_b\,E_W \Big( E_W (s_2(h^0) (1+3\,x_{h^0}-3\,x_t) + 
4\,s_1 (h^0) (1+2\,x_{h^0}))
\nonumber \\ &+& 
m_t\,(s_2 (h^0)+2\,s_1(h^0)\,(2\,x_{h^0}+x_t)) \Big) \Big)
\nonumber \\ &+&
m_W^2 \Big( m_t (s_2(h^0)\,(-1+(x_{h^0}-x_t)^2)+4\,s_1 (h^0)\,(1+x_t+x_{h^0}) 
+2\,E_b\,(-2\,s_1 (h^0)\,(1+2\,x_{h^0}+x_t) 
\nonumber \\ &+& 
s_2 (h^0)\,(-1+x_{h^0}-x_t)\, (2\, x_t-1))
\nonumber \\ &-&
2\,E_W\,(s_2(h^0)\,(x_{h^0}-x_t)\,(1-x_t+x_{h^0})+ s_1 (h^0)\,(4+2\,x_{h^0}\,
(2+2\,x_{h^0}-x_t)+2\,x_t\, (3-x_t))) \Big)
\nonumber \\ &+&  
\frac{8}{m_W^2}\,\Big( E_b\,E_W^2 \, (E_b (2\,E_W+m_t)\,s_2
(h^0)-4\,E_W\,m_t\,s_1 (h^0))  \Big) \Bigg) \, , \nonumber \\
\Phi^{A^0}&=& 4\,m_t\,s_1(A^0)\, |V_{tb}|^2\, \Bigg( 4 \Big(2\,E_W^2\,m_t
\,s_1(A^0)\, (x_t-2)-E_b^2\, (2\,E_W+m_t)\,s_2 (A^0)\,(1+x_t)
\nonumber \\ &+&
E_b\,E_W \Big( E_W (s_2(A^0) (1+3\,x_{A^0}-3\,x_t) + 
4\,s_1 (A^0) (1+2\,x_{A^0}))
\nonumber \\ &+& 
m_t\,(s_2 (A^0)+2\,s_1(A^0)\,(2\,x_{A^0}+x_t)) \Big) \Big)
\nonumber \\ &+&
m_W^2 \Big( -m_t (s_2(A^0)\,(-1+(x_{A^0}-x_t)^2)+4\,s_1{A^0}\,(1+x_t+x_{A^0}) 
\nonumber \\ &+&
2\,E_b\,(2\,s_1 (A^0)\,(1+2\,x_{A^0}+x_t)+ s_2 (A^0)\,(-1+x_t+2\,x_t^2+
x_{A^0} (1-2\,x_t))\,
\nonumber \\ &+&
2\,E_W\,\Big( s_2(A^0)\,(x_{A^0}-x_t)\,(1-x_t+x_{A^0})+ 
s_1 (A^0)\,(4+4\,x_{A^0}\,(1+x_{A^0}-2\,x_t)+2\,x_t\,(3-x_t)) \Big)\Big)
\nonumber \\ &+&  
\frac{8}{m_W^2}\,\Big( E_b\,E_W^2\, (E_b (2\,E_W+m_t)\,s_2 (A^0)-
4\,E_W\,m_t\,s_1 (A^0))  \Big) \Bigg) \, , 
\label{ACP2a}
\end{eqnarray}
with $\chi=e^{i\,\theta_{\chi}}\,|\chi|$, $\bar{\xi}^D_{N,bb}=
e^{i\,\theta_{bb}}\,|\bar{\xi}^D_{N,bb}|$. Notice that we do not present 
the functions $D(h^0)$ and $D(A^0)$ since their explicit expressions are 
long. Here the functions $s_1 (h^0 (A^0))$ and $s_2 (h^0 (A^0))$ are given 
in eq.(\ref{s1s2}). 
\section{Discussion}
This section is devoted to the analysis of the $BR$ and $A_{CP}$ of the decay 
$t\rightarrow b W h^0$ and $t\rightarrow b W A^0$ in the framework of model
III and beyond.  In our
numerical analysis we use the form of the coupling $\bar{\xi}^{U (D)}_{N,ij}$,
which  is defined as $\xi^{U(D)}_{N,ij}=\sqrt{\frac{4\,G_F}{\sqrt{2}}}\,
\bar{\xi}^{U(D)}_{N,ij}$.

Since the model III contains large number of free parameters such as Yukawa 
couplings, $\bar{\xi}^{U (D)}_{N, ij}$, the masses of new Higgs bosons, 
$H^{\pm}$, $h^0$ and $A^0$, we try to restrict them by using experimental 
measurements. In our calculations, we neglect all the Yukawa couplings except 
$\bar{\xi}^{U}_{N, tt}$ and $\bar{\xi}^{D}_{N, bb}$, due to their 
their light flavor contents.   In addition to this we neglect the off 
diagonal coupling $\bar{\xi}^{U}_{N, tc}$, since it is smaller compared to 
$\bar{\xi}^{U}_{N,tt}$ (see \cite{Alil1}). One of the most important
experimental measurement for the prediction of the constraint region for the
couplings $\bar{\xi}^{U}_{N, tt}$ and $\bar{\xi}^{D}_{N, bb}$ is the 
the CLEO measurement \cite{cleo2}
\begin{eqnarray}
BR (B\rightarrow X_s\gamma)= (3.15\pm 0.35\pm 0.32)\, 10^{-4} \,\, .
\label{br2}
\end{eqnarray}
and our procedure is to restrict the Wilson coefficient $C_7^{eff}$ which is 
the effective coefficient of the operator $O_7 = \frac{e}{16 \pi^2} 
\bar{s}_{\alpha} \sigma_{\mu \nu} (m_b R + m_s L) b_{\alpha} 
{\cal{F}}^{\mu \nu}$ (see \cite{Alil1} and references therein), in the 
region $0.257 \leq |C_7^{eff}| \leq 0.439$, where the upper and lower limits 
were calculated using eq. (\ref{br2}) and all possible uncertainities in the 
calculation of $C_7^{eff}$ \cite{Alil1}. In the case of the calculation of
$A_{CP}$, $\bar{\xi}^{D}_{N, bb}$ ($\bar{\xi}^{U}_{N, tt}$) is taken complex 
(real) and a new small complex parameter $\chi$, due to the physics beyond
model III, is introduced. In the following, we choose 
$|r_{tb}|=|\frac{\bar{\xi}_{N, tt}^{U}}{\bar{\xi}_{N, bb}^{D}}| 
<1$. Notice that, in figures,  the $BR$ and $A_{CP}$ are restricted in the 
region between solid (dashed) lines for $C_7^{eff} > 0$ ($C_7^{eff} < 0$). 
Here, there are two possible solutions for $C_7^{eff}$ due to the cases 
where $|r_{tb}|<1$ and $r_{tb}>1$. In the case of complex Yukawa couplings, 
only the solutions obeying $|r_{tb}|<1$ exist.

\begin{figure}[htb]
\vskip -3.0truein
\centering
\epsfxsize=6.8in
\leavevmode\epsffile{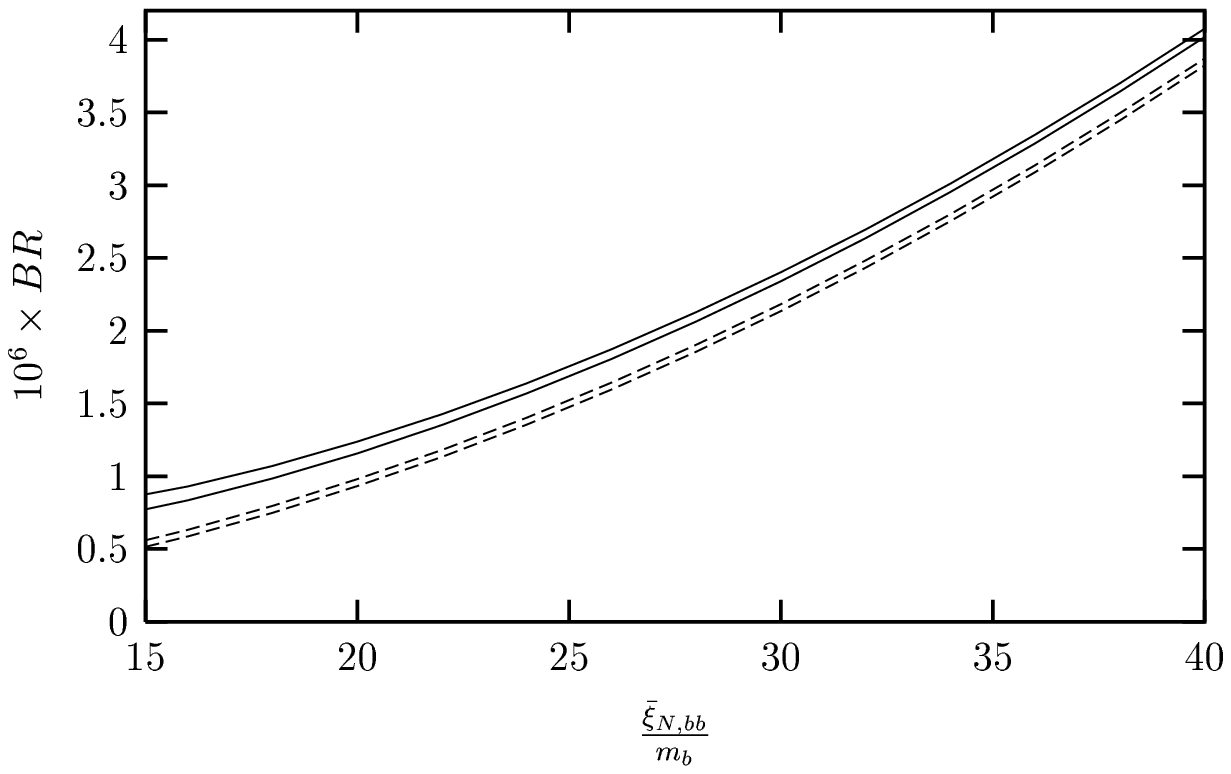}
\vskip -3.0truein
\caption[]{$BR (t\rightarrow b W h^0)$ as a function of 
$\frac{\bar{\xi}^{D}_{N, bb}}{m_b}$ for $m_{H^{\pm}}=400\, GeV$, 
$m_{h^0}=85\, GeV$. Here the $BR$ is restricted in the region bounded by 
solid lines for $C_7^{eff}>0$ and by dashed  lines for $C_7^{eff}<0$.}
\label{BRc1h0}
\end{figure}

In  Fig. \ref{BRc1h0}, we plot the $BR(t\rightarrow b W h^0)$ with respect 
to $\frac{\bar{\xi}^{D}_{N, bb}}{m_b}$ for $m_{H^{\pm}}=400\, GeV$, 
$m_{h^0}=85\, GeV$.  As shown in this figure, the $BR$ is at the 
order of the magnitude of $10^{-6}$ and it increases with the increasing 
values of the $\frac{\bar{\xi}^{D}_{N, bb}}{m_b}$. Its magnitude (the 
restriction region ) is larger (broader) for $C_7^{eff} > 0$ compared to 
the one for $C_7^{eff} < 0$.

\begin{figure}[htb]
\vskip -3.0truein
\centering
\epsfxsize=6.8in
\leavevmode\epsffile{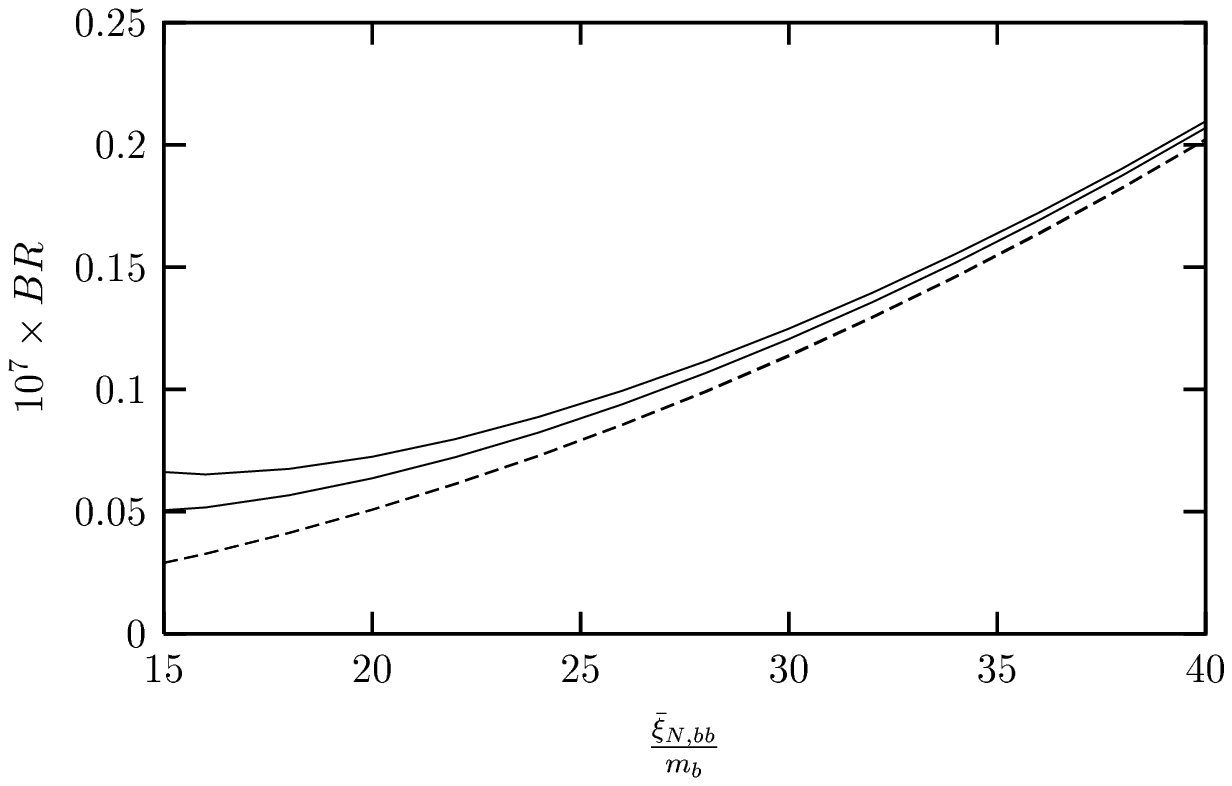}
\vskip -3.0truein
\caption[]{The same as Fig. \ref{BRc1h0} but for the decay 
$BR (t\rightarrow b W  A^0)$.}
\label{BRc1A0}
\end{figure}

Fig. \ref{BRc1A0} is devoted to the same dependence of the 
$BR(t\rightarrow b W A^0)$ for $m_{H^{\pm}}=400\, GeV$, 
$m_{A^0}=90\, GeV$.  For this process the $BR$ is at the order of the 
magnitude of $10^{-8}$, almost 2 order smaller compared to the 
$BR(t\rightarrow b W h^0)$. It increases with the increasing values 
of the $\frac{\bar{\xi}^{D}_{N, bb}}{m_b}$ and its magnitude  (the 
restriction region) is larger (broader) for $C_7^{eff} > 0$  compared to 
the one for $C_7^{eff} < 0$. Furthermore, the restriction region is
sensitive to the parameter $\frac{\bar{\xi}^{D}_{N, bb}}{m_b}$ and for 
$C_7^{eff} < 0$, upper and lower bounds almost coincide.  

\begin{figure}[htb]
\vskip -3.0truein
\centering
 \epsfxsize=6.8in
\leavevmode\epsffile{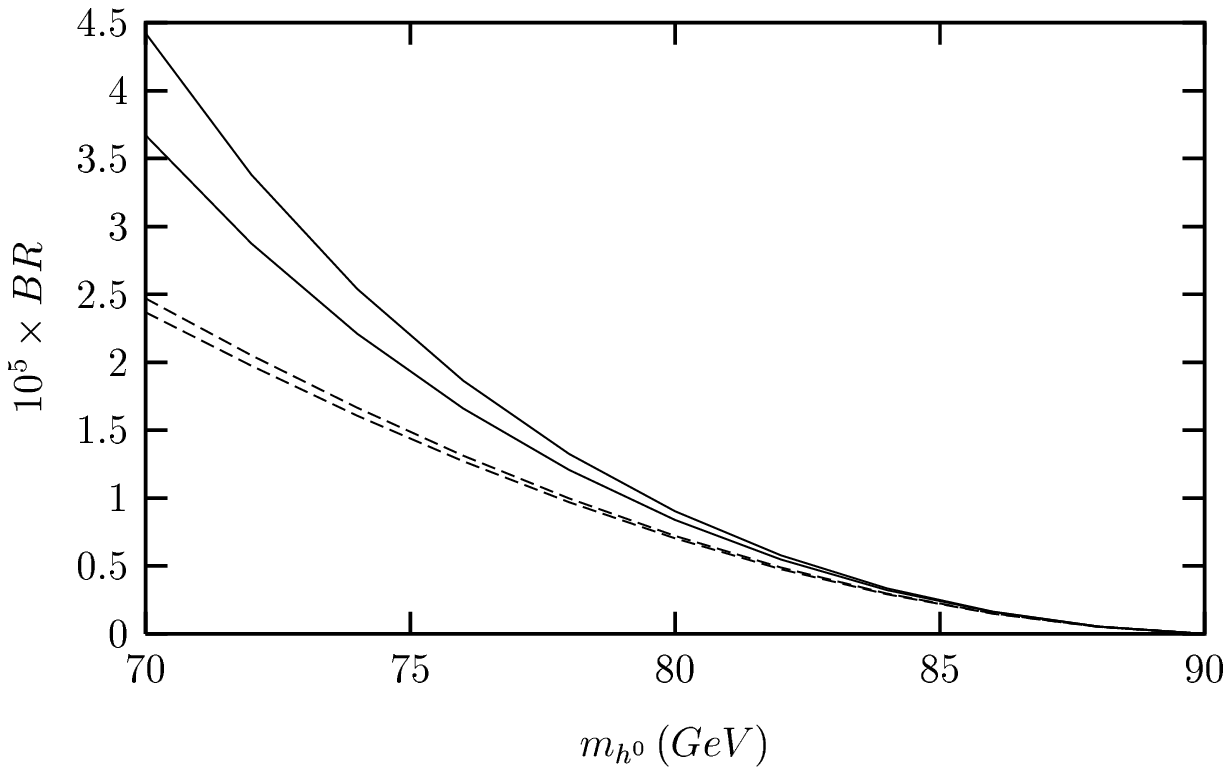}
\vskip -3.0truein
\caption[]{$BR (t\rightarrow b W h^0)$ as a function of $m_{h^0}$ for
$\bar{\xi}^{D}_{N, bb}=30\, m_b$, $m_{H^{\pm}}=400\, GeV$. 
Here the $BR$ is restricted in the region bounded by 
solid lines for $C_7^{eff}>0$ and by dashed  lines for $C_7^{eff}<0$.}
\label{BRmh0}
\end{figure}

\begin{figure}[htb]
\vskip -3.0truein
\centering
\epsfxsize=6.8in
\leavevmode\epsffile{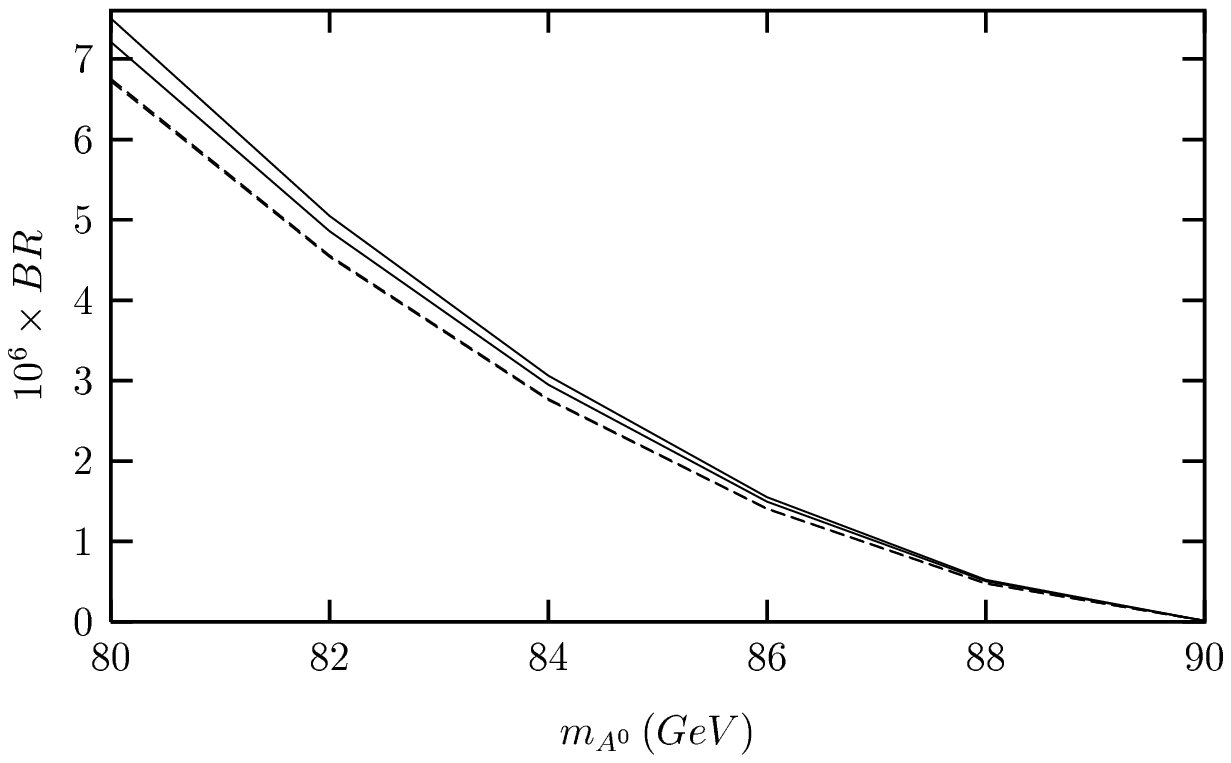}
\vskip -3.0truein
\caption[]{The same as Fig. \ref{BRmh0} but for the decay 
$BR (t\rightarrow b W  A^0)$.}
\label{BRmA0}
\end{figure}

Fig. \ref{BRmh0} (\ref{BRmA0}) represents $BR(t\rightarrow b W h^0 (A^0))$ 
with respect to $m_{h^0} (m_{A^0})$ for $m_{H^{\pm}}=400\, GeV$ and 
$\bar{\xi}^{D}_{N, bb}=30\,m_b$. Here the $BR$ increases with the 
decreasing values of $m_{h^0} (m_{A^0})$. This can give a powerfull 
information about the lower limit of the mass value $m_{h^0} 
(m_{A^0})$ with the help of the possible future experimental measurement of 
the process under consideration. Notice that with the increasing values of   
$m_{h^0} (m_{A^0})$ the restriction regions for $C_7^{eff} > 0$ and 
$C_7^{eff} < 0$ become narrower and coincide.

Now, we would like to analyse the CP asymmetry $A_{CP}$ of the decay
$t\rightarrow b W h^0 (A^0)$. To obtain a nonzero $A_{CP}$ we take the
coupling $\bar{\xi}^{D}_{N, bb}$ complex and introduce a new complex
parameter $\chi$ due to the physics beyond the model III (see section II).

\begin{figure}[htb]
\vskip -3.0truein
\centering
\epsfxsize=6.8in
\leavevmode\epsffile{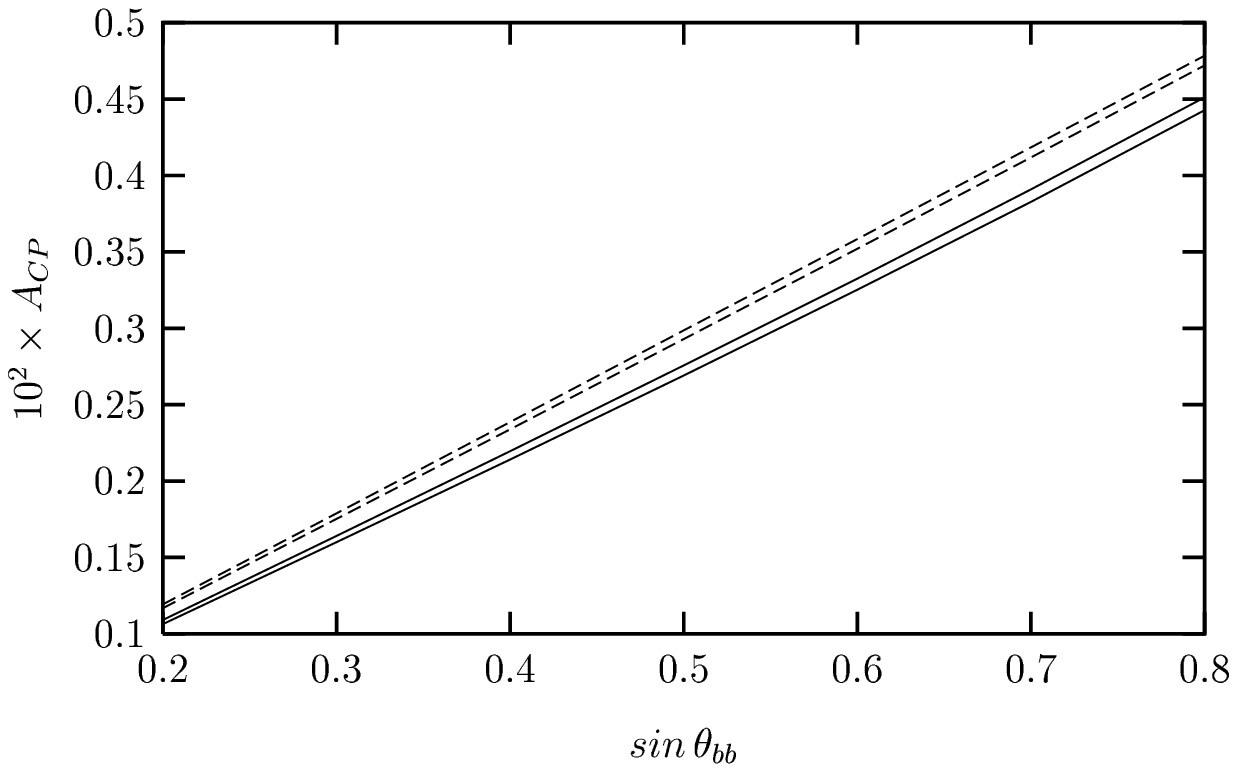}
\vskip -3.0truein
\caption[]{$A_{CP} (t\rightarrow b W h^0)$ as a function of $sin\theta_{bb}$ 
for $|\bar{\xi}^{D}_{N, bb}|=30\, m_b$, $m_{H^{\pm}}=400\, GeV$,
$|\chi|=10^{-2}$, $sin\theta_{\chi}=0.5$. Here the $A_{CP}$ is restricted in 
the region bounded by solid lines for $C_7^{eff}>0$ and by dashed lines for 
$C_7^{eff}<0$.}
\label{ACPsinbbh0}
\end{figure}

\begin{figure}[htb]
\vskip -3.0truein
\centering
\epsfxsize=6.8in
\leavevmode\epsffile{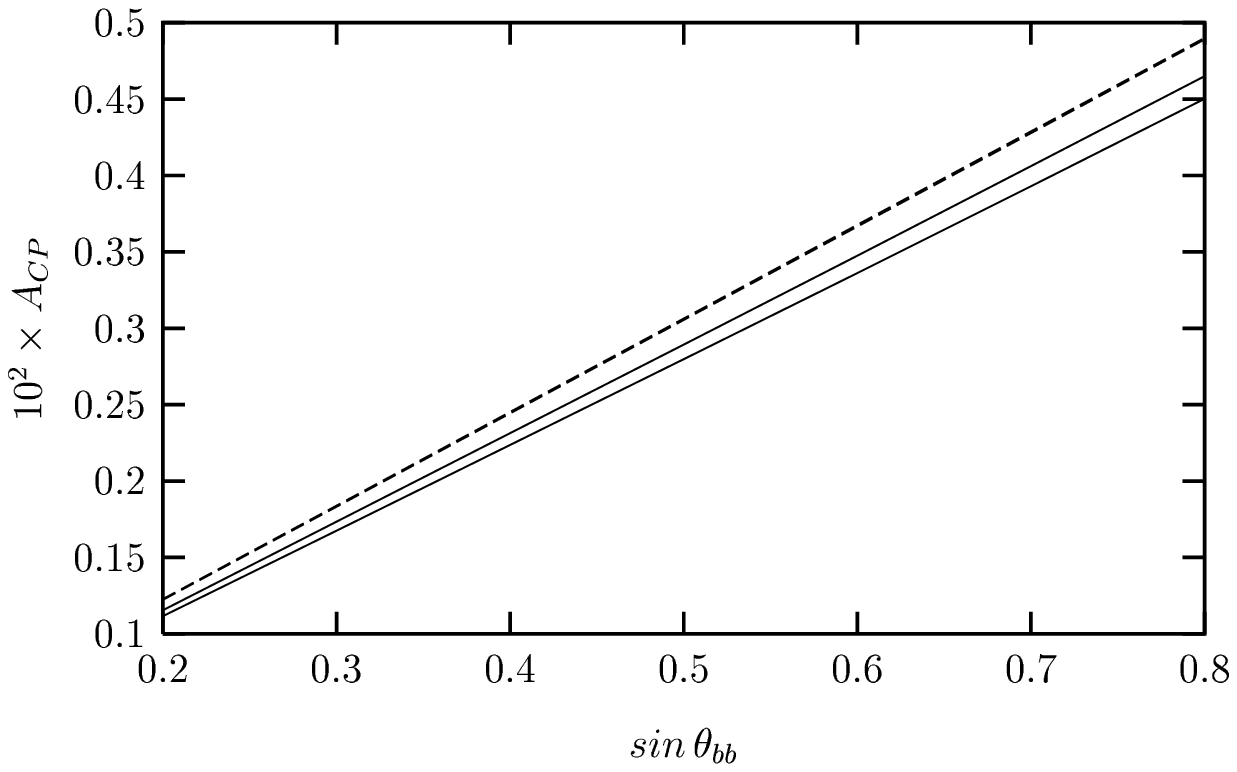}
\vskip -3.0truein
\caption[]{The same as Fig. \ref{ACPsinbbh0} but for the decay 
$A_{CP} (t\rightarrow b W  A^0)$.}
\label{ACPsinbbA0}
\end{figure}

\begin{figure}[htb]
\vskip -3.0truein
\centering
\epsfxsize=6.8in
\leavevmode\epsffile{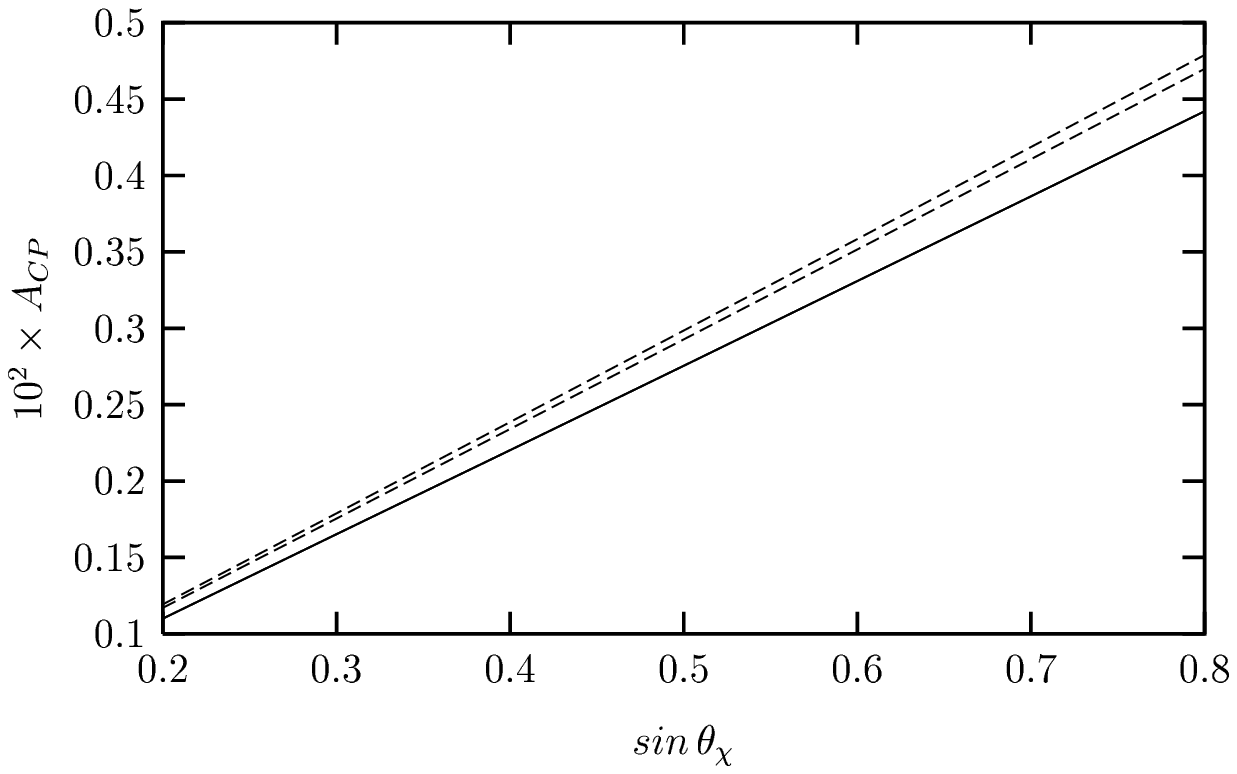}
\vskip -3.0truein
\caption[]{$A_{CP} (t\rightarrow b W h^0)$ as a function of
$sin\theta_{\chi}$ for $|\bar{\xi}^{D}_{N, bb}|=30\, m_b$, $m_{H^{\pm}}=
400\, GeV$, $|\chi|=10^{-2}$, $sin\theta_{bb}=0.5$. Here the $A_{CP}$ is 
restricted in the region bounded by solid lines for $C_7^{eff}>0$ and by 
dashed lines for $C_7^{eff}<0$.}
\label{ACPsinkaph0}
\end{figure}

\begin{figure}[htb]
\vskip -3.0truein
\centering
\epsfxsize=6.8in
\leavevmode\epsffile{ 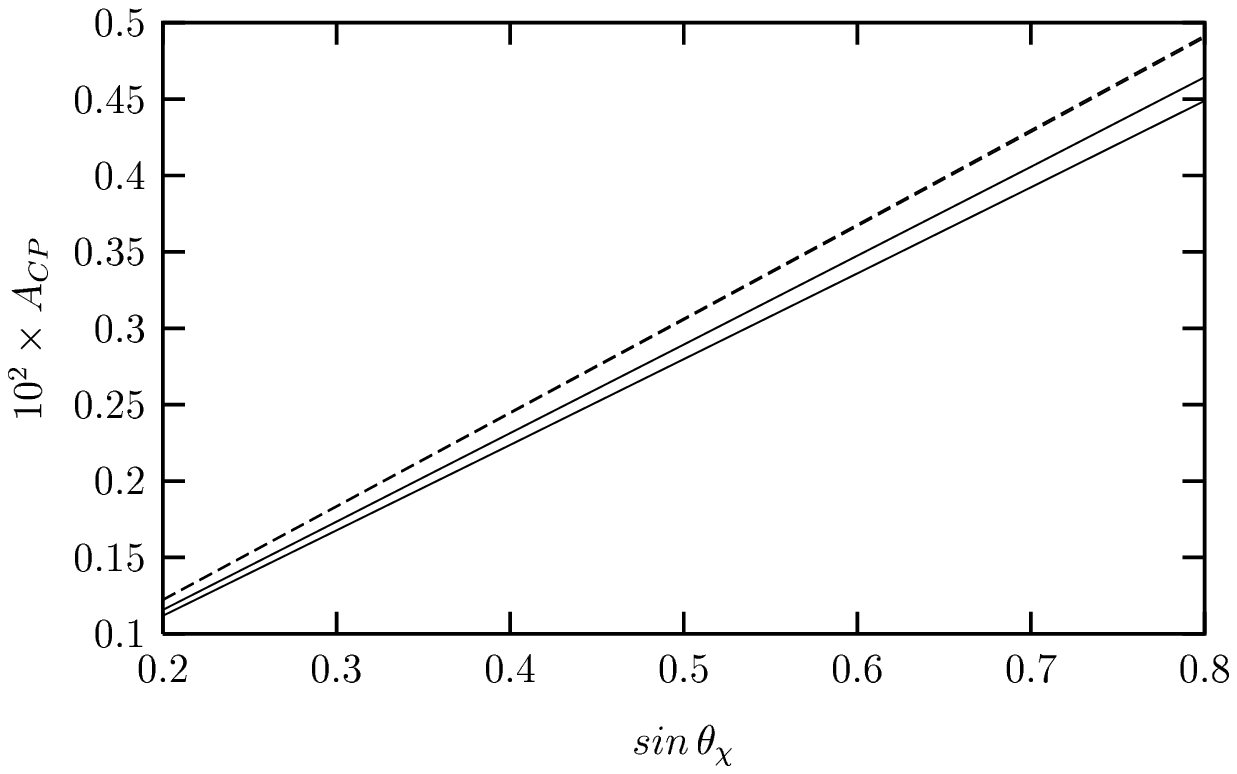}
\vskip -3.0truein
\caption[]{The same as Fig. \ref{ACPsinkaph0} but for the decay 
$A_{CP} (t\rightarrow b W  A^0)$.}
\label{ACPsinkapA0}
\end{figure}

In Fig \ref{ACPsinbbh0} (\ref{ACPsinbbA0}) we present the 
$sin\,\theta_{bb}$ dependence of $A_{CP}(t\rightarrow b W h^0 (A^0))$ 
for $|\bar{\xi}^{D}_{N, bb}|=30\,m_b$, $|\chi|=10^{-2}$, the intermediate 
value of $sin\theta_{\chi}=0.5$ and $m_{h^0}=85\, GeV \, (m_{A^0}=90\, GeV)$.
$A_{CP}$ is at the order of the magnitude of $10^{-3}-10^{-2}$ and slightly
larger for $C_7^{eff} < 0$ compared to the one for $C_7^{eff} > 0$.  
Fig \ref{ACPsinkaph0} (\ref{ACPsinkapA0}) represents the 
$sin\,\theta_{\chi}$ dependence of $A_{CP}(t\rightarrow b W h^0 (A^0))$ 
for $|\bar{\xi}^{D}_{N, bb}|=30\,m_b$, $|\chi|=10^{-2}$, the 
intermediate value of $sin\theta_{bb}=0.5$ and $m_{h^0}=85\, GeV \, 
(m_{A^0}=90\, GeV)$. The behavior of $A_{CP}$ is similar to the one obtained 
in Fig. \ref{ACPsinbbh0} (\ref{ACPsinbbA0}). As shown in these figures, 
$A_{CP}$ is a mesurable quantity, which gives strong clues about the possible 
physics beyond the SM. 

Now we will summarize our results:

\begin{itemize}

\item The $BR$ of the process $t\rightarrow b W h^0 (A^0)$ is at the 
order of $10^{-6}$ ($10^{-8}$) in the model III and it can be measured 
in the future experiments. This ensures a crucial test for the new physics 
beyond the SM. 

\item The $BR$ is sensitive to $\bar{\xi}^{D}_{N, bb}$ and the mass value 
$m_{h^0}$ ($m_{A^0}$). This is important in the prediction of the lower 
limit of the mass $m_{h^0} (m_{A^0})$ with the possible future 
experimental measurement of the process under consideration.

\item $A_{CP}$ is at the order of the magnitude of $10^{-2}$ for the 
intermeditate values of $sin\,\theta_{bb}$ and and $sin\,\theta_{\chi}$. 
This a mesurable quantity, which gives a strong clue  about the possible 
physics beyond the SM and also more beyond. 

\end{itemize}

Therefore, the experimental investigation of the process $t\rightarrow b 
W h^0 (A^0)$ will be effective for understanding the physics beyond the 
SM.
\section{Acknowledgement}
This work was supported by Turkish Academy of Sciences (TUBA/GEBIP).

\newpage
\newpage
\end{document}